\documentclass[runningheads]{llncs}
\usepackage[T1]{fontenc}
\usepackage{amsmath} 
\usepackage{algorithm}
\usepackage{algpseudocode}
\usepackage{amssymb}
\usepackage{subcaption}
\usepackage{booktabs}

\usepackage{graphicx}
\usepackage[usenames,dvipsnames]{xcolor}
\newif\ifshowcomments
\showcommentstrue 
\ifshowcomments

\def\Gc#1{\textcolor{blue}{\textbf{\texttt{ \footnotesize [#1]}}}} 
 
\def\Cc#1{\textcolor{green}{\textbf{\texttt{\textit{\footnotesize [#1]}}}}} 
\def\Pc#1{\textcolor{orange}{P:\textbf{#1}}}
\else

\def\Gc#1{} 

\def\Cc#1{} 

\def\Pc#1{} 
\fi
\begin{document}
\title{Detection of Emerging Infectious Diseases in Lung CT based on Spatial Anomaly Patterns}
\titlerunning{Detection of Emerging Infectious Diseases}
%
 \author{Branko Mitic\inst{1}\orcidID{0009-0006-7838-2370} \and
 Philipp Seeb\"ock\inst{1}\orcidID{0000-0001-5512-5810} \and
 Jennifer Straub\inst{1}\orcidID{0000-0002-2576-1717} \and
 Helmut Prosch\inst{2}\orcidID{0000-0002-6119-6364} \and
 Georg Langs\inst{1}\orcidID{0000-0002-5536-6873}}

%
\authorrunning{B. Mitic et al.}
%

\institute{$^1$ Computational Imaging Research Lab, and $^2$ Division of General and Pediatric Radiology, Department of Biomedical Imaging and Image-guided Therapy, Medical University of Vienna, Spitalgasse 23, 1090 Vienna, Austria  \\\email{branko.mitic@meduniwien.ac.at, georg.langs@meduniwien.ac.at} \\
\url{https://www.cir.meduniwien.ac.at}
\\}

\maketitle              

\begin{abstract}

Fast detection of emerging diseases is important for containing their spread and treating patients effectively. Local anomalies are relevant, but often novel diseases involve familiar disease patterns in new spatial distributions. Therefore, established local anomaly detection approaches may fail to identify them as new. Here, we present a novel approach to detect the emergence of new disease phenotypes exhibiting distinct patterns of the spatial distribution of lesions. We first identify anomalies in lung CT data, and then compare their distribution in a continually acquired new patient cohorts with historic patient population observed over a long prior period. We evaluate how accumulated evidence collected in the stream of patients is able to detect the onset of an emerging disease. In a gram-matrix based representation derived from the intermediate layers of a three-dimensional convolutional neural network, newly emerging clusters indicate emerging diseases. 

\end{abstract}


\section{Introduction}

Clinical imaging has been crucial in identifying and characterizing outbreaks of respiratory diseases that became pandemics, including SARS, MERS, H1N1, and COVID-19 \cite{Wilder-Smith2020-rz}. For COVID-19, computed tomography (CT) provided more specific features than symptoms like fever, cough, and muscle pain \cite{chassagnon2021ai}. CT revealed distinct ground-glass opacities (GGOs) and patterns in unusual areas such as basal and peripheral regions \cite{ding2020chest,pan2020time,shi2020radiological,wang2020imaging,zhou2020imaging}. Imaging was vital for early detection and characterization of COVID-19 cases and remains essential in ongoing patient management \cite{fang2021covid}.

Fast identification of emerging infectious diseases is important for controlling their spread and ensuring effective treatment of patients. However, accumulating sufficient evidence observed by individual clinicians in the form of clinical manifestations in temporal and regional clusters is not trivial. It relies on continued surveillance and coordinated communication. Here, we propose an algorithm to detect emerging disease phenotypes in a stream of patient imaging data. 

\begin{figure}[t]
  \centering
  \includegraphics[width=1\linewidth]{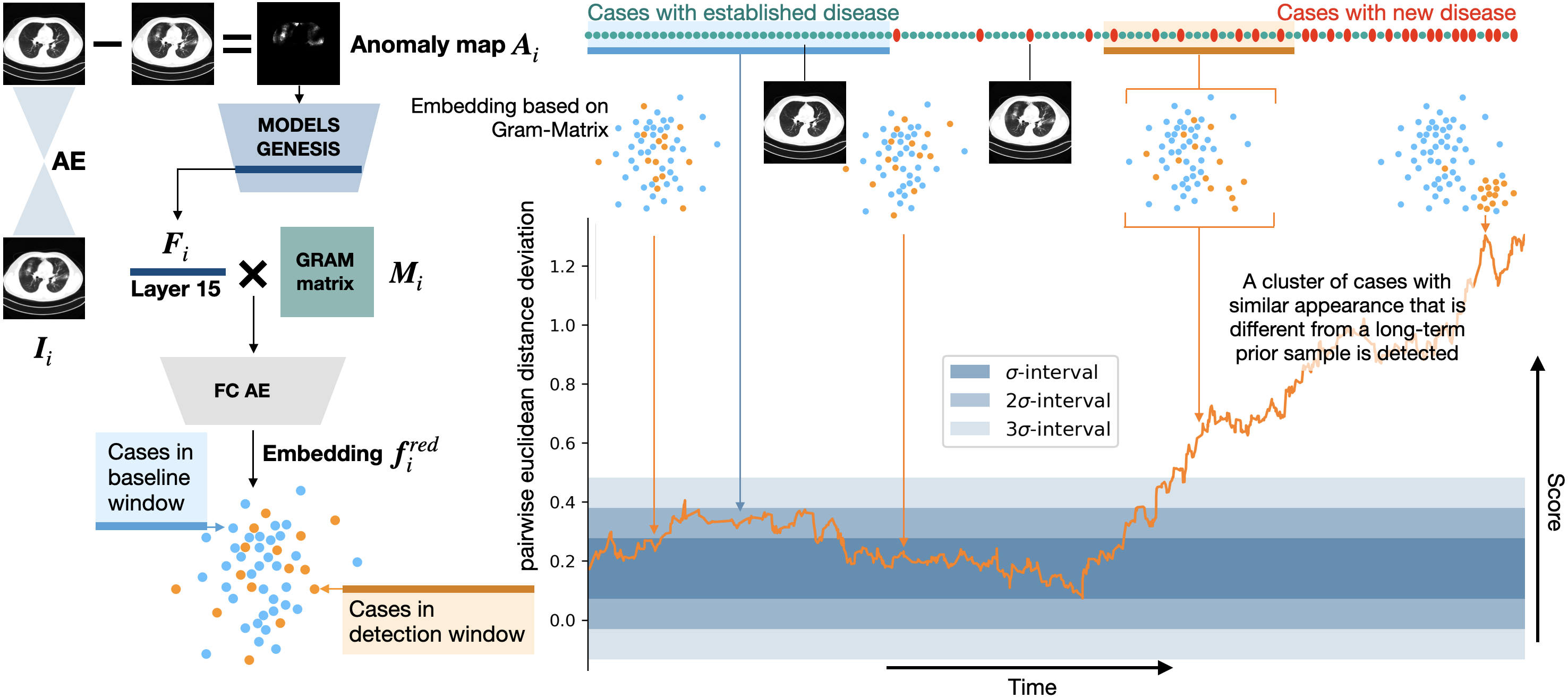}
  \caption{To detect a novel disease phenotype, we compute anomaly maps and extract features using a pre-trained 3D U-Net \cite{ronneberger2015u,zhou2021models}. We then calculate the gram-matrix and multiply it with the intermediate feature representation to obtain a disease appearance descriptor. An autoencoder-based dimensionality reduction provides the final case embeddings. This distribution helps identify new clusters of cases that differ from previous patient populations while being similar to each other. 
}  
\label{fig: main_new}
\end{figure}

\paragraph{Related work} In anomaly detection, notable successes have been achieved within the medical domain, particularly through the application of generative networks in retinal imaging \cite{schlegl2017unsupervised,schlegl2019f,seebock2018unsupervised}. Transformer-, and diffusion models have shown promise in anomaly detection in brain MRIs \cite{pinaya2021unsupervised,pinaya2022fast}. Conversely, there has been a limited number of publications focusing on anomaly detection and localization in CT-scans and chest X-rays related to lung conditions \cite{yadav2021lung,nakao2021unsupervised,kim20223d}. Outbreak detection for epidemiology directly through imaging features is an understudied field: To the best of our knowledge, there is only one publication related to monitoring novel infectious diseases in lung X-rays~\cite{chharia2022deep}. Similarly to \cite{chharia2022deep}, our research aims at detecting emerging diseases in the context of pandemic prevention. In contrast to \cite{chharia2022deep} we focus on the domain of CT-imaging data, and assess patients cohorts, instead of classifying individual examples. 

\paragraph{Contribution} 
We propose an approach (Fig.~\ref{fig: main_new}) to detect a newly emerging disease in CT examinations of a continual stream of patients. We address the challenge of novel diseases exhibiting known local disease patterns by a representation that captures three-dimensional spatial distribution patterns of anomalies using a gram-matrix of an intermediate representational network layer. First, we use anomaly detection in the lung CT data to segment disease patterns present in the lung, independent from their specific class. Then, instead of identifying individual cases that are different from a baseline distribution, we detect clusters of cases presenting with image phenotypes that are at the same time similar and new. As those accumulate a score comparing the feature representations serves as an indicator of a newly emerging disease. Experimental results demonstrate that this approach identifies new diseases, even if the disease load does not change. Quantitative analysis describes the relationship between the number of necessary days until detection and the speed of the disease spread, with a more rapid spread enabling detection faster with fewer affected patients.

\section{Method}

The method operates on a stream of CT-volumes $\boldsymbol{I}_i$, with unknown diagnoses $y_i \in \{0,1\}$. $0$ denotes the set of all diagnoses present in the population before the onset of a new disease, and $1$ is a different disease that occurs for the first time at a specific time point (\textit{onset}) and whose frequency increases according to the reproduction number $R$. We aim at detecting the presence of a new disease in the stream of patients, after as few cases as possible. The method consists of three steps. \textbf{(1)} First, we detect anomalies as an anomaly map $\boldsymbol{A}_i$ in each image  $\boldsymbol{I}_i$ identifying regions visually affected by disease without constraining detection to specific known lesion types. \textbf{(2)} To capture the spatial lesion distribution, we create a feature representation $\boldsymbol{f}^l_i$ of the three-dimensional anomaly map using a pre-trained convolutional neural network (CNN) \cite{zhou2021models}, calculate the gram-matrix $\boldsymbol{M}^l_i$ \cite{gatys2015neural} at an intermediate layer $l$ of this CNN, and multiply it with the intermediate feature representation $\boldsymbol{F}^l_i$ to fuse local and global features to obtain $\boldsymbol{f}^l_i$. We reduce the dimensionality of the resulting vector by means of an autoencoder. \textbf{(3)} To detect a new disease in the stream of feature representations, we compare the distribution of the present patient population to the distribution of a prior baseline population in this embedding space. A score identifying newly forming dense clusters that are different from the prior patients serves as an indicator for the presence of a new disease. 

\subsection{Detecting disease patterns in images $\boldsymbol{I}_i\rightarrow \boldsymbol{A}_i$}\label{sec: 2.1}

We train an adversarial convolutional autoencoder (AE) \cite{isola2017image} on healthy samples to extract anomalous lung areas from CT-slices. The convolutional AE learns a mapping from healthy lungs augmented with a wide range of artificially simulated disease lesions to the corresponding healthy lung $G: \boldsymbol{H}^g_{i}\rightarrow \boldsymbol{H}_i$ \cite{goodfellow2014generative,isola2017image}. The training objective and implementation details are analogous to \cite{isola2017image}. The generator architecture is based on the U-Net framework, featuring an encoder-decoder structure with skip connections \cite{ronneberger2015u}. The discriminator is based on a modified VGG16 model \cite{simonyan2014very}. After training, we obtain the anomaly map of a new image by subtracting $\boldsymbol{H}_i$ from the input $\boldsymbol{H}^g_{i}$, yielding $ \boldsymbol{A}_i = \boldsymbol{H}^g_{i}-\boldsymbol{H}_i $. 

\subsection{Describing disease appearance $\boldsymbol{A}_i\mapsto \boldsymbol{f}^{red}_i$}\label{sec: 2.2}

\paragraph{Gram-matrix-based feature extraction}
For feature extraction, we first process anomaly maps $\boldsymbol{A}_i$ using truncated Models Genesis \cite{zhou2021models}. Next we compute the gram-matrix $\boldsymbol{M}^l$, defined as the matrix product between all vectorized feature maps in layer $l$; $\boldsymbol{M}^l= \boldsymbol{F}^l_{j}\cdot (\boldsymbol{F}^l_{j})^T$. This is called a \textit{style representation} \cite{gatys2015neural}, and in our case captures the appearance of spatial anomaly distribution patters. Finally, we calculate the matrix product of Models Genesis output and its corresponding gram-matrix, $\boldsymbol{f}_i=\boldsymbol{M}^l_i \cdot \boldsymbol{F}^l_i$ to combine local- and global characteristics, in a single feature vector.

\paragraph{Dimensionality reduction}

To reduce the dimensionality of $\boldsymbol{f}_i$, a fully connected autoencoder is trained on feature vectors of lungs augmented with artificially simulated disease patterns as input. We use standard mean squared error reconstruction loss for training the model. The encoder of the  model is then used to obtain the final embedding $\boldsymbol{f}^{red}_i$. Together these steps map a CT-volume to a feature vector $\boldsymbol{I}_i \mapsto \boldsymbol{f}^{red}_i$.

\subsection{Detecting a novel disease $\boldsymbol{f}^{red}_i\mapsto d_{dev}, s_{kde}$}
\label{subsec:method_detectionOfNovelDisease}

To detect the onset of a novel disease in a continuous stream of imaging data, we compute the mean pairwise euclidean distance of all feature vectors observed within two different time windows in order to quantify the compactness of an emerging cluster: a baseline time window $\boldsymbol{W}^b$ yielding average distance $d_b$ and a detection window $\boldsymbol{W}^d$ yielding $d_d$. The mean pairwise euclidean distance deviation is the absolute distance between these two values and is denoted $d_{dev} = \vert d_d - d_b \vert$. The baseline and detection window contain the same number of patients $w_{s}$. The onset of a new disease is detected if $d_{dev} > s * \sigma_w$, with $s$ controlling the detection sensitivity and the average pairwise euclidean distance standard deviation $\sigma_w$ for known diseases of a specific window size. Using a sliding window approach, we continuously analyse incoming cases in the patient stream and finally calculate the average required days until onset detection $n_{ped}$. Additionally, we determine whether the emerging cluster is part of the existing distribution or is actually new. To this end, we fit a Gaussian kernel density estimate to the baseline window and then calculate the mean log-likelihood of all samples in the detection window under the baseline model. This score $s_{kde}$ is used in the same way as $d_{dev}$ to find the average required days until onset detection $n_{kde}$. The overall proposed approach maps individual images $\boldsymbol{I}_i$ to corresponding feature representations $\boldsymbol{f}^{red}_i$, and the set of $w_{s}$ most recent patients to (1) a distance based score and (2) kernel density estimate score that together constitute a marker for a newly emerging disease in this population.

\section{Experiments and Evaluation}

We performed experiments on both simulated and real-world imaging data. By simulating various disease patterns in lung CT images, we evaluated our algorithm's ability to detect newly emerging diseases. Additionally, we tested our method on real-world data, including cases of community-acquired pneumonia and COVID-19.

\paragraph{Simulated data: simulating lung CT affected by different diseases}. The healthy data used in this work is a derivative of \cite{zhang2020clinically}. For evaluation we simulate GGOs in five different configurations $D$, with \textit{fragmentation} and \textit{spatial distribution within the lung} as varied parameters (Table~\ref{tab:data_table}). These configurations represent five different simulated diseases, with $D_5$ used as new disease in our experiments. The implementation of the filament-like structures in the GGOs is achieved with an algorithm developed for simulating cosmological large-scale structure formation in the early Universe \cite{porqueres2020hierarchical}.

\paragraph{Real world data:} For disease onset experiments in data containing real diseases we use a subset of healthy, covid-infected and pneumonic volumes from \cite{zhang2020clinically}, the details are summarized in Table ~\ref{tab:data_table}.

\begin{figure}[t!]
  \centering
  \includegraphics[width=1\linewidth]{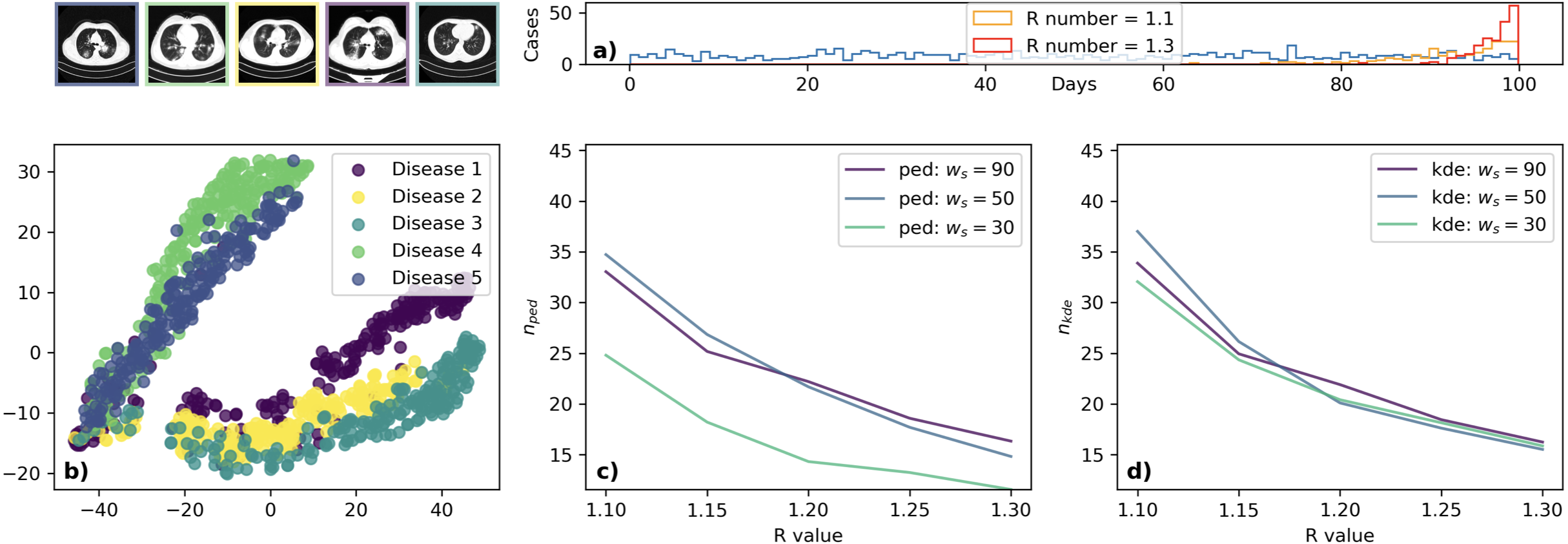}
  \caption{(a) The introduction of previously unseen diseases is shown for two different rates in orange and red and compared to the steady flow of known diseases. (b) t-SNE visualization of the embedding of a patient population with 5 different diseases. (c) and (d) The average days $n_{ped}$ and $n_{kde}$ necessary for positive detection as a function of reproduction number $R$ for three window sizes $w_s$.}
  \label{fig: combofig0}
\end{figure}

\begin{table}[t!]
\caption{Data overview, (a) evaluation dataset with five different simulated disease patterns ($D_1-D_5$), differing in \textit{fragmentation (high, low)}, and \textit{spatial distribution (front, back, random)}; (b) real world dataset for slice-wise anomaly extractor training and disease onset detection experiments.}
\begin{subtable}{0.5\linewidth}
    \centering
    \begin{tabular}{p{1cm}|p{1cm}p{1cm}p{1cm}p{1cm}p{1cm}}
       \textbf{(a)}  & D1 & D2 & D3 & D4 & D5 \\
        \hline
        \textbf{frag.} & low & low & high & high & low \\
        \textbf{loc.} & rand. & front & front & back & back \\
        \textbf{cases} & 200 & 200 & 200 & 200 & 200 \\
        &  &  &  &  &  \\
        \multicolumn{6}{c}{}\\

    \end{tabular}
\end{subtable}%
\begin{subtable}{0.5\linewidth}
    \centering
    \begin{tabular}{l|p{1.5cm}p{1cm}p{1cm}}
        \textbf{(b)} & healthy & cap & covid \\
        \hline
         cases & 128 & 128 & 128 \\
         train. & 1044 2D & & \\
         val. & 32 & 32 & \\
         test & 32 & 32 & \\
         \multicolumn{4}{c}{}\\
    \end{tabular}
\end{subtable}%
\label{tab:data_table}
\end{table}
\paragraph{Evaluating disease onset detection with different disease introduction rates}
To simulate the gradual onset of a new disease into a continual stream of known diseases, we sample timestamps $\in [0,100]$ in days as units from a uniform distribution for known diseases. The timestamps $\in [50,100]$ for new diseases are sampled from an exponential distribution with probability density function $f(x;\frac{1}{\gamma})=\frac{1}{\gamma}\exp (-\frac{x}{\gamma})$, where $\gamma$ denotes the scale parameter which is related to the reproduction number $R$ via $\gamma=\frac{1}{\ln (R)}$ (Fig.~\ref{fig: combofig0} (b)). After sorting all samples according to their timestamp, we apply our method. We use $w_s=\{30,50,90\}$, Models Genesis layer $l=5$ and detection sensitivity $s=2.576$ corresponding to an interval including $99\%$ of the baseline data in our experiments for both simulated data and real data. In the first experiment, we vary the $R$ value in five steps from $1.1$ to $1.3$: $[1.1,1.15,1.2,1.25,1.3]$. This affects the introduction rate of novel diseases (Fig.~\ref{fig: combofig0}(b)). For each of the five $R$ values, we create 100 time series by randomly sampling timestamps for all evaluation cases. We evaluate the average required number of days until disease detection $n_{ped}$ and $n_{kde}$ based on mean pairwise Euclidean distance deviation $d_{dev}$ and kernel density score $s_{kde}$ separately.

\paragraph{Evaluating the pixel level anomaly extractor}
We validate the quality of the anomaly extractor by testing its ability to separate healthy and infected data in a subset from \cite{zhang2020clinically} (Table~\ref{tab:data_table}), where the test- and validation sets contain 32 healthy and 32 pneumonic patient volumes each. 
For this we extract anomaly maps $\boldsymbol{A}_i$ from cases in the validation set and compute the pixel sum per case. The optimal pixel sum threshold for classification directly results from the pixel sum distributions, i.e. every case from the test set is classified as unhealthy if its pixel sum falls beyond the threshold value determined with the validation set. Finally, we calculate the percentage of correctly identified unhealthy cases as a measure of classification accuracy and consequently anomaly extractor quality.

\begin{figure}[t!]
  \centering
  \includegraphics[width=1\linewidth]{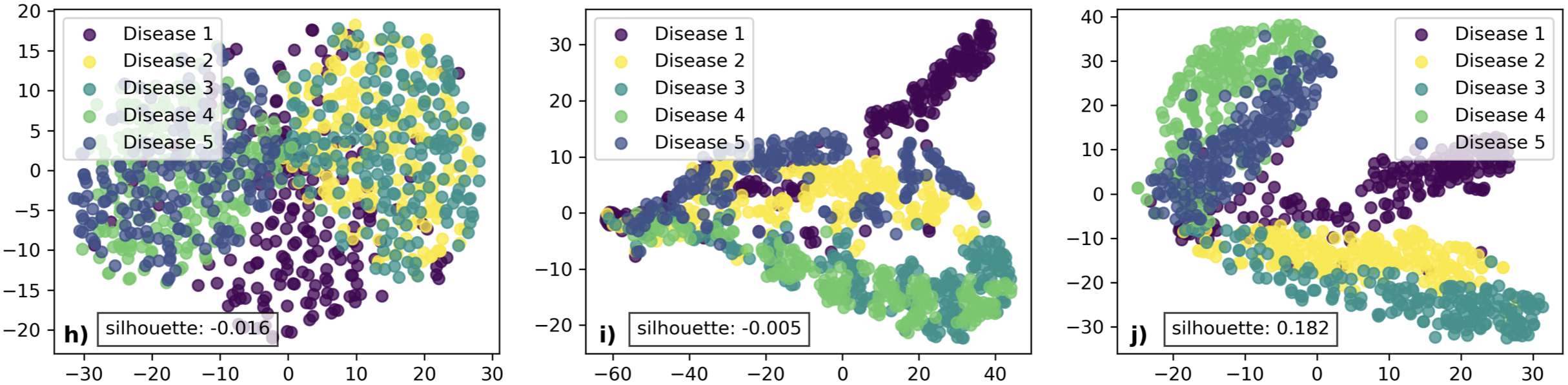}
  \caption{Embedding spaces capturing the population with 5 simulated diseases illustrating the separation of different diseases: (h) intermediate layer output features; (i) gram-matrix;  (j) proposed feature representation.}
  \label{fig: combofig2}
\end{figure}

\paragraph{Ablation study of the patient level feature representation}
To evaluate the impact of feature representation on the separation of different diseases in the embedding space we assess the separation of the 5 diseases in (1) only the VGG-layer $\boldsymbol{F}_i$ of the anomaly maps and 2) only  the gram-matrix $\boldsymbol{M}_i$ of the anomaly maps, and (3) the proposed feature representation in Fig.~\ref{fig: combofig2} h), i) and j). Ultimately, we report separation of the diseases using the Silhouette value, and show a qualitative visualization of the embedding space using t-distributed Stochastic Neighbor Embedding (t-SNE) \cite{van2008visualizing}.

\paragraph{Implementation details of the convolutional autoencoder}
The generator model is based on Isola~et~al.~\cite{isola2017image}. For the discriminator we utilize a VGG16 model \cite{simonyan2014very} pre-trained on ImageNet \cite{deng2009imagenet}, truncated at block five and layers prior to block five set to non-trainable. Finally, we add a convolutional layer with a single 4 × 4 filter followed by a sigmoid activation function. We use the ADAM optimizer \cite{kingma2014adam} with a $\beta_1$ value of $0.5$, a learning rate of $2\cdot 10^{-4}$ and train the model for 12 epochs on the training set (Table~\ref{tab:data_table}(b)).

\paragraph{Fully connected autoencoder implementation details} 
Denoting a dense layer with a ReLu activation function and $k$ units as $Dk$ the encoder-decoder architecture is defined as $D4096 - D2048 - D1024 - D512$ for the encoder and $D1024 - D2048 - D4096 - D131072$ for the decoder. The activation function in the last layer of the decoder is a sigmoid function. We train the model with SGD optimizer, a learning rate of $1$ and train for 100 epochs. For training we simulate 800 anomaly maps with disease patterns P1 to P4, purposely excluding the new disease P5. The same training procedure is repeated for real data.


\begin{figure}[t!]
  \centering
  \includegraphics[width=1\linewidth]{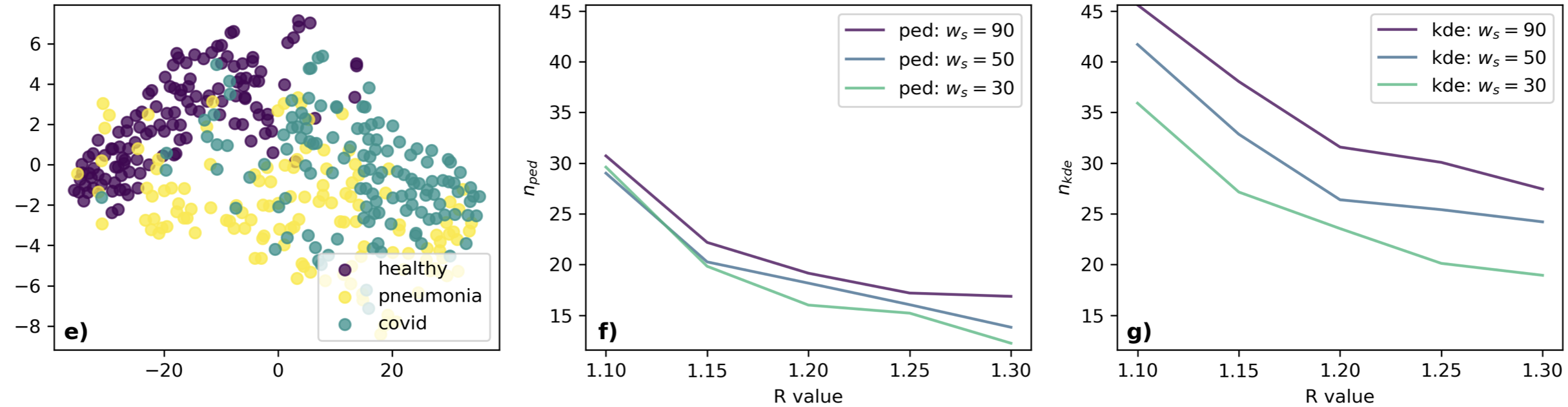}
  \caption{Analogous to Fig.~\ref{fig: combofig0}, the results for real data are in panels e) to g).}
  \label{fig: combofig1}
\end{figure}

\section{Results}
\begin{table}[t!]
\begin{center}
\caption{Results for pairwise euclidean distance based and kernel density score based detection in simulated and real data.}
\begin{tabular}{ r|llllll } 
  \hline

Simulated data & $R=1.10$ & $R=1.15$ & $R=1.20$ & $R=1.25$ & $R=1.30$ & FP/100 days \\ 
  \hline
ped $w_s=90$ & 33.04 & 25.17 & 22.20 & 18.59 & 16.33 & 8.63\\ 
ped $w_s=50$ & 34.72 & 26.84 & 21.69 & 17.70 & 14.82 & 13.16\\ 
ped $w_s=30$ & 24.80 & 18.20 & 14.32 & 13.24 & 11.57 & 7.84\\ 
kde $w_s=90$ & 33.88 & 24.94 & 21.90 & 18.45 & 16.24 & 9.51\\ 
kde $w_s=50$ & 37.01 & 26.15 & 20.10 & 17.61 & 15.52 & 9.44\\ 
kde $w_s=30$ & 32.05 & 24.34 & 20.41 & 18.14 & 15.86 & 16.76\\ 
  \hline
Real data & & & & & & \\ 
  \hline
ped $w_s=90$ & 45.60 & 38.04 & 31.58 & 30.08 & 27.45 & 1.93\\ 
ped $w_s=50$ & 41.70 & 32.85 & 26.39 & 25.41 & 24.21 & 2.59\\ 
ped $w_s=30$ & 35.91 & 27.15 & 23.55 & 20.12 & 18.94 & 2.66\\ 
kde $w_s=90$ & 30.71 & 22.19 & 19.15 & 17.19 & 16.87 & 0.92\\ 
kde $w_s=50$ & 29.00 & 20.26 & 18.17 & 16.04 & 13.82 & 3.10\\ 
kde $w_s=30$ & 29.60 & 19.83 & 16.00 & 15.21 & 12.25 & 3.36\\ 

\end{tabular}
\label{tab:results_table}
\end{center}
\end{table}

Different diseases are visible in different regions in the feature embedding for simulated- and real data in Fig.~\ref{fig: combofig0}b) and  Fig.~\ref{fig: combofig1} e) suggesting that our method successfully captures both global and local disease pattern differences. Fig.~\ref{fig: combofig0}c), d) and Fig.~\ref{fig: combofig1} f) and g) show the time to detection for different window sizes and the two different scores, with a fixed score threshold as defined above across the experiments. Not surprisingly, increasing R decreases the time to detection. Table\,\ref{tab:results_table} shows the corresponding values, together with the false positive rates. While $n_{ped}$ and $n_{kde}$ are comparable for simulated data, the kde-based detection outperforms the ped-based detection on real data across introduction rates, and window sizes.  When comparing different sizes $w_s$ of the detection window, results show a trend with window size $w_s=90$ leading to an increased number of required novel cases to detect the onset of the new disease. This decreased sensitivity, however, leads to higher specificity, i.e. the number of falsely triggered onset detections decreases with increased window size. The anomaly extractor, used to detect anomalies in the CT imaging data, achieves a classification accuracy of $84$\% with no false positives, demonstrating the effectiveness of our anomaly extractor. Results of the ablation study are shown in Fig.~\ref{fig: combofig2}h), i) and j). The t-SNE visualizations illustrate that the proposed feature representation offers better separation of the diseases compared to the two individual components quantified by the Silhouette values of the four feature variants, with the proposed features yielding the highest value of 0.182, compared to -0.005 for gram-matrix and -0.016 for intermediate layer as a basis for calculation. 

\section{Discussion and Conclusion}

We propose a new method to identify emerging diseases in a continuous stream of CT imaging data. Results demonstrate that our approach is able to detect novel diseases even in a scenario when their appearance differs only in subtly distinct spatial distributions of already known anomalies. This contributes to the ability to detect newly emerging diseases based on population level clinical imaging data, a largely unexplored problem. To facilitate varied experimental evaluation, we introduce a novel strategy to simulate realistic GGO disease patterns in lungs.
The approach is evaluated in a setting of simulated GGO lung anomaly patterns. In contrast to biases in real data where the detection of certain diseases might be possible with relatively simple metrics capturing e.g. the disease load only, the simulated setting offers two primary advantages: (1) full control over quantity and quality of the data and (2) certainty regarding class membership.
Results show a strong influence of the reproduction value $R$ on the number of required days for disease onset detection. We are convinced that this is a reasonable property of the model, as slowly emerging diseases require more time to form distinct clusters in the embedding space.
The advantage of smaller window sizes across most experiments suggests that new clusters can be detected with few examples, and dilution of the set of novel disease cases in larger windows is detrimental to detection sensitivity. This will be subject of future analysis. 

This work serves as a proof of concept towards the unsupervised detection of novel emerging diseases in real clinical routine imaging data.

\textbf{Acknowledgments} This work was funded by the Austrian Science Fund (FWF, P 35189-B - ONSET), the European Commission (No. 101080302 - AI-POD), and the Vienna Science and Technology Fund (WWTF, PREDICTOME [10.47379/LS20065]) and is part of AIX-COVNET and ZODIAC Zoonotic Disease Integrated Action of the IAEA/UN.


\bibliographystyle{splncs04}

\end{document}